\documentclass[11pt,a4paper]{article}

\usepackage{jheppub}
\usepackage[normalem]{ulem}

\usepackage[utf8]{inputenc}
\usepackage{graphicx}
\usepackage{epsfig}
\usepackage{subfigure}
\usepackage{color}
\usepackage{comment}
\usepackage{amsmath}
\usepackage{mathtools}
\usepackage{tabularx}
\usepackage{tabu}
\usepackage{hhline}
\usepackage{multirow}
\usepackage{bm}
\usepackage{tensor}
\usepackage{ulem}

\newcommand{\be}{\begin{equation}}
\newcommand{\ee}{\end{equation}}

\providecommand{\f}[2]{\frac{{#1}}{{#2}}}
\newcommand{\Lagr}{\mathcal{L}}

\def\Nbub{\langle {\cal N}\rangle}

\title{The effective Higgs potential and vacuum decay in Starobinsky inflation}

\author[a]{Andreas Mantziris} 
\author[b,c]{\!, Tommi Markkanen}
\author[a]{and Arttu Rajantie}

\affiliation[a]{Department of Physics, Imperial College London, London, SW7 2AZ, United Kingdom}
\affiliation[b]{Solita Oy, Eteläesplanadi 8, 00130 Helsinki, Finland}
\affiliation[c]{Helsinki Institute of Physics, P.O. Box 64, FIN-00014 University of Helsinki, Finland}

\emailAdd{a.mantziris18@imperial.ac.uk}
\emailAdd{tommi.markkanen@solita.fi}
\emailAdd{a.rajantie@imperial.ac.uk}

\abstract{
Based on the current experimental data, the Standard Model predicts that the current vacuum state of the Universe is metastable, leading to a non-zero rate of vacuum decay through nucleation of bubbles of true vacuum. Our existence implies that there cannot have been any such bubble nucleation events anywhere in our whole past lightcone. We consider a minimal scenario of the Standard Model together with Starobinsky inflation, using three-loop renormalization group improved Higgs effective potential with one-loop curvature corrections. We show that the survival of the vacuum state through inflation places a lower bound $\xi\gtrsim 0.1$ on the non-minimal Higgs curvature coupling, the last unknown parameter of the Standard Model. This bound is significantly stronger than in single field inflation models with no Higgs-inflaton coupling. It is also sensitive to the details of the dynamics at the end of inflation, and therefore it can be improved with a more detailed study of that period.
}

\begin{document}

\maketitle

\section{Introduction}

It is possible that the SM itself could describe physics up to the high energy scales of the very early Universe: This is due to the experimentally determined mass of the Higgs boson~\cite{Chatrchyan:2012ufa, Aad:2012tfa} lying in a range within which the Higgs self-interaction does not diverge below the Planck scale \cite{Degrassi:2012ry,Buttazzo:2013uya,Bednyakov:2015sca}. On the other hand, the current experimental data suggest that the SM Higgs vacuum appears to be metastable due to the existence of a second, lower minimum state in the Higgs potential, where its decay would induce the nucleation of true vacuum bubbles that expand rapidly and devour their surroundings, the implications of which has been the topic of many investigations \cite{Hung:1979dn,Sher:1993mf,Casas:1996aq,Isidori:2001bm,Ellis:2009tp,EliasMiro:2011aa,Lebedev:2012zw,Branchina:2013jra}.

It is often stated that because the bubble nucleation rate in the current Universe is extremely low, this metastability is not in conflict with observation. However, this is not enough, because the Universe must have also survived its cosmological evolution until the present day. Therefore the question of vacuum stability has to be looked at in the cosmological context. To be precise, there cannot have been any bubble nucleation events within our whole past lightcone~\cite{Espinosa:2007qp,Markkanen:2018pdo}, including the current late-time Universe, the whole thermal history of the Universe, and the inflationary era. Therefore vacuum stability and survival can be used to place non-trivial new constraints on cosmological scenarios and fundamental theories.

The cosmological implications of the Higgs vacuum metastability have been studied in a variety of scenarios in recent years and reviewed in \cite{Markkanen:2018pdo}. See the Refs. \cite{Espinosa:2007qp,Espinosa:2018mfn,Lebedev:2012sy,Kobakhidze:2013tn,Fairbairn:2014zia,Hook:2014uia,Kamada:2014ufa,Espinosa:2015qea,Kearney:2015vba,East:2016anr,Enqvist:2014bua,Bhattacharya:2014gva,Herranen:2014cua,Czerwinska:2015xwa,Rajantie:2016hkj,Czerwinska:2016fky,Rajantie:2017ajw,Markkanen:2018bfx,Rodriguez-Roman:2018swn,Rusak:2018kel,Jain:2019wxo,Hertzberg:2019prp,Lalak:2020dao,Adshead:2020ijf,Fumagalli:2019ohr} focusing on inflation, the Refs. \cite{Herranen:2015ima,Kohri:2016wof,Gross:2015bea,Ema:2016kpf,Enqvist:2016mqj,Ema:2017loe,Postma:2017hbk,Figueroa:2017slm,Croon:2019dfw,Kost:2021rbi} for studies discussing the reheating phase and the Refs. \cite{Kawasaki:2016ijp,Burda:2015isa,Burda:2016mou,Espinosa:2017sgp,Kohri:2017ybt,Espinosa:2018euj,Franciolini:2018ebs,Cline:2018ebc,Espinosa:2018eve,Hook:2019zxa,Hayashi:2020ocn} discussing cosmological signatures and other observable consequences. For recent works addressing aspects of vacuum decay in de Sitter space see the Refs. \cite{Cruz:2022ext,Camargo-Molina:2022paw,Camargo-Molina:2022ord,Vicentini:2022pra,Maeso:2021xvl}.

In particular, vacuum stability allows us to constrain the non-minimal gravitational coupling $\xi$ of the Higgs field~\cite{Espinosa:2007qp,Herranen:2014cua,Herranen:2015ima}. This parameter is required by renormalizability of the theory in curved spacetime~\cite{Chernikov:1968zm,Tagirov:1972vv,Callan:1970ze}, but it is practically impossible to measure its value experimentally in the current almost flat Universe. In contrast, vacuum stability during inflation~\cite{Espinosa:2007qp,Herranen:2014cua,Markkanen:2018bfx,Mantziris:2020rzh} and after inflation~\cite{Herranen:2015ima,Figueroa:2017slm,Li:2022wuf} requires it to lie within a narrow range around the conformal value $\xi=1/6$. For recent work on other cosmological implications of non-minimal couplings see the Refs. \cite{Figueroa:2021iwm,Clery:2022wib,Lebedev:2022ljz}.

In our previous work Ref.~\cite{Mantziris:2020rzh}, we studied the electroweak (EW) vacuum instability of the Standard Model (SM) Higgs field in the context of cosmological inflation to obtain lower bounds on the Higgs-curvature coupling. We made use of an RG improved Higgs potential calculated on a curved background and also took into account the time dependence by numerically solving the evolution equations. We considered three one-parameter models of inflation, quadratic, quartic and Starobinsky-like power law inflation and managed to constrain the coupling $\xi \gtrsim 0.06$, while denoting the effects of the uncertainty in the top quark mass. 

One of the inflationary potentials we investigated in Ref.~\cite{Mantziris:2020rzh} corresponded to Starobinsky inflation~\cite{Starobinsky:1980te,Vilenkin:1985md}, in which the inflaton field arises from a scalar metric degree of freedom when the action has a quadratic $R^2$ curvature term. It can be thought of as the minimal inflationary model because it does not require the introduction of any new fields, and it is also compatible with observations. However, in the presence of other scalar fields, specifically the Higgs field, it will give rise to derivative couplings between the fields, which we have not included in Ref.~\cite{Mantziris:2020rzh}. These terms have been investigated in the setting of mixed Higgs-Starobinsky inflation \cite{Ema:2017rqn, He:2018gyf}, and recently also to study vacuum stability during reheating in Starobinsky inflation~\cite{Li:2022wuf}.

In this work, we extend the analysis of Ref.~\cite{Mantziris:2020rzh} to study the Higgs vacuum metastability in Starobinsky inflation, in order to bound $\xi$ from below. In Section 2, we review the Starobinsky inflation model and derive the action for a scalar spectator field in the Einstein frame. In Section 3, we calculate the Higgs effective potential by including renormalization group improvement at three loops, one-loop spacetime curvature corrections, and contributions arising from the conformal transformation to the Einstein frame. In Section 4, we obtain the expression for the expected number of bubbles in our past lightcone, and in Section 5 we derive a lower bound on $\xi$ by demanding that this number must be less than one. Finally, we present and discuss our conclusions in Section 6.

\section{Starobinsky Inflation}
\subsection{Inflaton potential}

In General Relativity (GR), gravity is described by the Einstein-Hilbert (EH) action
\begin{equation}
    \label{eq:EinsteinHilbert}
    S = \int d^4 x \sqrt{-g} \frac{M_P^2}{2}R,
\end{equation}
where $R$ is the Ricci scalar and $M_P = \left( 8 \pi G \right)^{-1/2} \approx 2.435 \times 10^{18}$ GeV is the reduced Planck mass. However, one would generally expect that the true gravitational action includes other terms, due to for example gravitational induced renormalization group running or the conformal anomaly (see for example Refs.~\cite{Markkanen:2013nwa,Markkanen:2018bfx}). In the simplest such extensions, known as $f(R)$ theories, the action is given by a different function $f(R)$ of the Ricci scalar,
\begin{align}
    S = \int d^4 x \sqrt{-g_J} \frac{M_P^2}{2}f(R_J) \, ,
    \label{eq:action_R_R2}
\end{align}
where $J$ denotes the original metric $g_{J \mu \nu}$ in which the theory is defined, and which is known as the Jordan frame. As we will see, it is possible to carry out a change of variables to a different metric $g_{\mu\nu}$ in which the action has the Einstein-Hilbert form (\ref{eq:EinsteinHilbert}), and which is known as the Einstein frame. In this paper, we will focus on the simplest $f(R)$ theory, which modifies the EH action by a quadratic term,
\begin{align}
    f(R_J) = R_J + \frac{R_J^2}{6 M^2 M_P^2} \, ,
    \label{eq:f(R)}
\end{align}
where $M$ is a small dimensionless parameter. This is equivalent to the inflationary model proposed by Starobinsky in 1980~\cite{Starobinsky:1980te,Vilenkin:1985md}, which is known as Starobinsky or $R^2$ inflation.

In order to describe the physics of this theory, it is convenient to carry out the transformation to Einstein frame. To do that, we first remove the quadratic term $R_J^2$ by introducing an auxiliary scalaron field $s$
with the action
\begin{align}
    S = \int d^4 x \sqrt{-g_J} \left[ \frac{M_P^2}{2} \left( 1 + \frac{s}{3 M^2 M_P^2} \right) R_J - \frac{s^2}{12 M^2}\right] \, ,
    \label{eq:auxs}
\end{align}
whose classical equation of motion is  $s=R_J$ on-shell and thus reproduces the action from (\ref{eq:action_R_R2}) and (\ref{eq:f(R)}). Now, the action can be turned into the EH form (\ref{eq:EinsteinHilbert}) by a conformal transformation
\begin{align}
    g_{\mu \nu} = \Omega^2 g_{J \mu \nu} \,
    \label{eq:metricCT}
\end{align}    
where
\begin{align}
    \Omega^2 = 1 + \frac{s}{3 M^2 M_P^2}.
    \label{eq:OmegaCT0}
\end{align}
Because the conformal transformation is not a coordinate transformation, it changes the Ricci scalar. As a result, the Ricci scalars $R_J$ and $R$ corresponding to the Jordan and Einstein frames, respectively, are related by the equation
\begin{align}
    R_J = \Omega^2 \left[ R - 3 \Box \mathrm{ln}\Omega^2 + \frac{3}{2} g^{\mu \nu} \partial_{\mu}\mathrm{ln}\Omega^2 \partial_{\nu}\mathrm{ln}\Omega^2 \right] \, ,
    \label{eq:RicciCT}
\end{align}
which gives rise to new terms in the action.

To write this in a more convenient form, we carry out a new change of variables and introduce a scalar field $\phi$, which we refer to as the inflaton, through
\begin{align}
    \Omega^2 = 1 + \frac{s}{3 M^2 M_P^2} = e^{\sqrt{\frac{2}{3}} \frac{\phi}{M_P}} \,.
    \label{eq:OmegaCT}
\end{align}
In terms of $\phi$, the relation between the two Ricci scalars (\ref{eq:RicciCT}) becomes
\begin{align}
    R_J = \Omega^2 \left[ R - \frac{\sqrt{6}}{M_P}\Box\phi + \frac{1}{M_P^2} g^{\mu \nu} \partial_{\mu}\phi \partial_{\nu}\phi \right] \, .
    \label{eq:RicciCT2}
\end{align}
In the Einstein frame, and written in terms of $\phi$, the action (\ref{eq:RicciCT}) is therefore
\begin{equation}
    S = \int d^4x \sqrt{-g} \left[ \frac{M_P^2}{2} R + \frac{1}{2} \partial_{\mu}\phi \partial^{\mu}\phi - V_{\rm I}(\phi) \right] \, ,
    \label{eq:EinsteinS}
\end{equation}
where we have omitted the $\Box\phi$ term because it is a total derivative, and we have introduced the inflaton potential
\begin{equation}
    V_{\rm I}(\phi) = \frac{3 M^2 M_P^4}{4} \left(1 - e^{-\sqrt{\frac{2}{3}} \frac{\phi}{M_P}} \right)^2 \, .
    \label{eq:V_I}
\end{equation}

In conclusion we can see that the modified gravity theory defined by Eq.~(\ref{eq:f(R)}) can be equivalently viewed as a Einstein gravity with an additional scalar field $\phi$ with potential $V_I(\phi)$. When $\phi\gtrsim M_P$, the potential satisfies the slow roll conditions, and therefore leads to inflation. Because it does not require an introduction of any additional fields by hand, only a small and well-justified modification of the gravitational action, it can be viewed as the minimal model of inflation. It is also in great agreement with observational constraints \cite{planck:2018jri}, thus making it one of the most promising models to describe the inflationary epoch. The value of the single free parameter $M$ can be determined from the observed amplitude of the cosmic microwave background temperature anisotropies to be $M=1.1 \times 10^{-5}$ \cite{Liddle:2000cg, planck:2018jri}.

\subsection{Non-minimally coupled scalar spectator field}

In this paper, since we are interested in the evolution of the Higgs field, and therefore we also need to understand how the scalar field action appears in the Einstein frame. Let us, therefore, consider a spectator scalar doublet field $\Phi$ non-minimally coupled to spacetime curvature with potential $V_{\Phi}$ in the Jordan frame, whose action reads
\begin{align}
    S =  \int d^4x \sqrt{-g_J} \bigg[ \frac{M_P^2}{2} \left( 1 - \frac{\xi \Phi^\dagger\Phi}{M_P^2} \right) R_J + \frac{1}{12 M^2}R_J^2 +  \frac{1}{2} g_J^{\mu \nu} (\partial_{\mu} \Phi^\dagger) (\partial_{\nu} \Phi)  - V_{\Phi}  \bigg] \, ,
\end{align}   
where $\xi = 1/6$ corresponds to the conformal point in our convention. We assume that $\Phi$ is a quantum field in the classical background metric, and we ignore the backreaction of $\Phi$ on the metric.

As in Eq.~(\ref{eq:auxs}), we remove the quadratic curvature term by introducing the auxiliary field $s$, so that the action becomes
\begin{align}
     S = \int d^4x \sqrt{-g_J} \bigg[ \frac{M_P^2}{2} \left( 1 + \frac{s}{3 M^2 M_P^2} \right) R_J - \frac{1}{12 M^2}s^2  + \frac{1}{2} g_J^{\mu \nu} (\partial_{\mu} \Phi^\dagger) (\partial_{\nu} \Phi) - \frac{\xi}{2} \Phi^\dagger\Phi R_J - V_{\Phi} \bigg] \, .
    \label{eq:action0}
\end{align}    
We then use the same conformal transformation (\ref{eq:OmegaCT}), to write the action in the Einstein frame as
\begin{multline}
    S = \int d^4x \sqrt{-g} \left[ \frac{M_P^2}{2} R + \frac{1}{2} \partial_{\mu}\phi \partial^{\mu}\phi \left(1 - \frac{\xi e^{-\sqrt{\frac{2}{3}} \frac{\phi}{M_P}} \Phi^\dagger\Phi }{M_P^2} \right) \right. \\ 
    \left. + \sqrt{\frac{3}{2}}\frac{\xi e^{-\sqrt{\frac{2}{3}} \frac{\phi}{M_P}} \Phi^\dagger\Phi}{M_P} \Box \phi + \frac{e^{-\sqrt{\frac{2}{3}} \frac{\phi}{M_P}}}{2} \partial_{\mu} \Phi^\dagger \partial^{\mu} \Phi - U(\phi, \Phi) \right] \, ,
 \label{eq:action1}
\end{multline}
where we have grouped the potential terms together as
\begin{align}
    U (\phi, \Phi) = V_{\rm I} (\phi) + \frac{\xi}{2} \frac{\Phi^\dagger\Phi R}{e^{\sqrt{\frac{2}{3}} \frac{\phi}{M_P}}} + \frac{V_{\Phi}}{e^{2 \sqrt{\frac{2}{3}} \frac{\phi}{M_P}}} \, .
    \label{eq:potential1}
\end{align}

To canonically normalise the scalar field $\Phi$, we rescale it with the field redefinition
\begin{align}
    \Phi = e^{\frac{1}{2} \sqrt{\frac{2}{3}} \frac{\phi}{M_P}} \Tilde{\Phi} \,,
    \label{eq:field-redef}
\end{align}
which turns the action to
\begin{multline}
    S = \int d^4x \sqrt{-g} \left[ \frac{M_P^2}{2}R + \frac{1}{2}\partial_{\mu} \phi \partial^{\mu} \phi \left( 1 + \left( -\xi + \frac{1}{6} \right)  \frac{\Tilde{\Phi}^\dagger\Tilde{\Phi}}{M_P^2} \right) \right.
    \\ + \sqrt{\frac{3}{2}}\left( -\xi +\frac{1}{6} \right) \frac{\partial^{\mu}  \phi}{M_P} 
    \partial_\mu\left(\Tilde\Phi^\dagger \Tilde{\Phi}\right)
    \left. + \frac{1}{2} \partial_{\mu} \Tilde{\Phi}^\dagger \partial^{\mu}  \Tilde{\Phi}  - \tilde{U}(\phi,\Tilde\Phi) \right] \, ,
 \label{eq:action3}
\end{multline}
with
\begin{equation}
    \tilde{U}(\phi,\tilde\Phi)
    =V_{\rm I} (\phi) + \frac{\xi}{2} \Tilde{\Phi}^\dagger\tilde{\Phi} R + e^{-2\sqrt{\frac{2}{3}} \frac{\phi}{M_P}}
    V_{\Phi} \left(e^{\frac{1}{2} \sqrt{\frac{2}{3}} \frac{\phi}{M_P}} \Tilde{\Phi}\right) \, .
\end{equation}

In the case of a renormalizable tree-level potential
\begin{equation}
    V_\Phi(\Phi)=\frac{1}{2}m^2\Phi^\dagger\Phi+ \frac{1}{4}\lambda\left(\Phi^\dagger\Phi\right)^2 \, ,
\end{equation}
 we obtain
\begin{equation}
\label{eq:Utilde}
    \tilde{U}(\phi,\tilde\Phi)
    =V_{\rm I} (\phi) + \frac{\xi}{2} \Tilde{\Phi}^\dagger\tilde{\Phi} R 
    + \frac{1}{2}e^{-\sqrt{\frac{2}{3}} \frac{\phi}{M_P}}m^2 \Tilde\Phi^\dagger\Tilde\Phi+ \frac{1}{4}\lambda\left(\Tilde\Phi^\dagger\Tilde\Phi\right)^2 \, ,
\end{equation}
which shows that in the Einstein frame, and expressed in terms of $\Tilde\Phi$, the potential has the same form as in the original Jordan frame, but with a scaled mass term.

More generally, we also observe that the transformation from Jordan to Einstein frame has given rise to two non-renormalizable coupling terms between $\phi$ and $\Phi$. We will discuss these terms and their effect on the scalar field dynamics further in Section~\ref{sec:timedep}. However, we can also note that if the non-minimal coupling is conformal, i.e., $\xi=1/6$, or if $\phi$ is constant so that $\partial_\mu\phi=0$, these terms vanish, and therefore the scalar field action has its standard form. The latter is a good approximation during inflation, and we make use of it in Section~\ref{sec:effective_masses}, when we compute quantum corrections to the effective Higgs potential.

\section{Effective Higgs potential} \label{sec:3}

\subsection{Renormalization group improved effective potential in curved spacetime}
\label{sec:RGI}

Vacuum metastability in the Standard Model is a quantum effect, which only appears at one loop order in perturbation theory. Therefore, in order to describe it, it is essential to compute the effective Higgs potential. In practice, we use the renormalization group improved (RGI) effective potential using three-loop beta functions~\cite{Chetyrkin:2012rz}, including curvature corrections at one-loop order~\cite{Markkanen:2018bfx}, in the approximation in which the spacetime has constant curvature $R$ and the inflaton field $\phi$ is constant so that $\partial_\mu\phi=0$. However, before including the effects of the $R^2$ term, we first summarise the calculation in conventional Einstein gravity, i.e. without the inflaton field $\phi$~\cite{Markkanen:2018bfx}.

The Standard Model particle content can be written as
\begin{equation}
{\cal L}_{\rm SM} ={\cal L}_{\rm YM} + {\cal L}_{\rm F} + {\cal L}_{\Phi}+{\cal L}_{GF}+{\cal L}_{GH} \,,
\label{eq:lag}
\end{equation}
where the first three terms in eq.~(\ref{eq:lag}) come from the gauge fields, the fermions and the Higgs doublet $\Phi$, respectively. The `GF' and `GH' correspond to the gauge fixing and ghost Lagrangians, respectively. Similarly as in Ref.~\cite{Mantziris:2020rzh}, we will be using the $\zeta_i =1$ choice for the gauge fixings throughout this calculation. Here we show only the steps for the Higgs contribution, but the complete derivation can be found in Ref.~\cite{Markkanen:2018bfx} (see also a shortened derivation in Ref.~\cite{Markkanen:2018pdo}).

The Higgs piece reads
\begin{align}
{\cal L}_{\Phi} = \left(D_\mu \Phi \right)^\dagger \left(D^\mu \Phi \right) -m^2 \Phi^\dagger \Phi-\xi R\Phi^\dagger \Phi- \lambda (\Phi^\dagger \Phi)^2\label{eq:higg1}
\,,
\end{align}
where $m^2<0$ is the Higgs mass parameter, $\xi$ is the non-minimal Higgs curvature coupling, and $\lambda$ is the Higgs self-coupling. The covariant derivative is
\begin{align}
D_\mu = \nabla_\mu - ig \tau^a A^a_\mu - ig'Y A_\mu;\qquad \tau^a=\sigma^a/2\, ,
\end{align}
where $\nabla_\mu$ contains the covariant connection for Einsteinian gravity, $g$ and $g'$ are the $SU(2)$ and $U(1)$ gauge couplings, $A^a_\mu$ and $A_\mu$ the gauge fields, $\tau$ and $Y$ the corresponding generators, and $\sigma^a$ are the Pauli matrices. Expressing the Higgs field $\Phi$ as\footnote{Here we use the same notation as in Ref.~\cite{Mantziris:2020rzh}, which differs slightly to that in Ref. ~\cite{Markkanen:2018pdo} ($\varphi\leftrightarrow h$)}
\begin{equation}
\Phi = \f{1}{\sqrt{2}}\left(\begin{array}{c} -i(\chi_1 - i \chi_2) \\ h+ (\chi_0 + i \chi_3)\end{array}\right)\,, 
\end{equation}
where $h\in\mathbb{R}$ is a constant classical mean field and $\chi_0$ and $\chi_i$, $i\in\{1,2,3\}$, are quantum fluctuations with zero expectation value, the scalar part of the Lagrangian becomes
\begin{equation}
{\cal L}_{\rm SM} = -\f{m^2}{2}h^2-\f{\lambda}{4}h^4
-\f{1}{2}\chi_0\left[\Box+m^2_h+\xi R\right]\chi_0
-\f{1}{2}\chi_i\left[\Box+m^2_\chi+\xi R\right]\chi_i
+\cdots
\,,
\end{equation}
where the effective masses are
\begin{equation}
m^2_h=m^2+3\lambda h^2\,,
\quad
m^2_\chi=m^2+\lambda h^2.
\label{eq:effm}
\end{equation}
In a similar fashion, one may derive the quadratic {terms} 
for all degrees of freedom in the SM.

Considering the entirety of the SM particle spectrum on curved spacetime, the effective Higgs potential was calculated to 1-loop order in Ref.~\cite{Markkanen:2018bfx}, where the dS approximation allows us to write in more compactly as
\begin{eqnarray}
V_{\rm H}(h, \mu, R)=\f{m^2}{2}h^2+\f{\xi}{2}Rh^2+\f{\lambda}{4}h^4+V_\Lambda-\kappa R+\frac{\alpha}{144} R^2 + \Delta V_{\rm loops} \,,
\label{eq:VeffSMdS}
 \end{eqnarray}
where we have suppressed all the implicit renormalization scale dependence and we identify the additional terms as the mass term ($1^{\rm st}$), the cosmological constant correction ($4^{\rm th}$), the correction to the EH term ($5^{\rm th}$), the radiatively generated curvature correction ($6^{\rm th}$), and the loop correction ($7^{\rm th}$) that sums over the SM degrees of freedom and reads
\begin{eqnarray}
     \Delta V_{\rm loops} (h, \mu, R) = \frac{1}{64\pi^2} \sum\limits_{i=1}^{31}\bigg\{ n_i\mathcal{M}_i^4 \bigg[\log\left(\frac{|\mathcal{M}_i^2 |}{\mu^2}\right) - d_i \bigg] +\frac{n'_i}{144}R^2\log\left(\frac{|\mathcal{M}_i^2 |}{\mu^2}\right)\bigg\}  \,. \,\,\,\,\,\,
     \label{eq:DeltaVloops}
\end{eqnarray}
The details regarding the form of (\ref{eq:DeltaVloops}) are contained in Section 5 of Ref.~\cite{Markkanen:2018bfx}. In the early universe, there is significant curvature to the spacetime continuum and therefore the SM effective masses receive curvature corrections, in addition to flat space contributions of the type as in eq.~(\ref{eq:effm}), which we denote with $\mathcal{M}_i$. In the context of Starobinsky inflation, the mass term of the Higgs field is negligible when compared to the high scales of the Hubble rate. Furthermore, setting $m = 0$ means also that the rest of the dimensionful couplings in the Lagrangian can be neglected due to the RG flow fixed point at $m = V_{\Lambda} = \kappa = 0$ \cite{Hardwick:2019uex}, which simplifies the Higgs potential (\ref{eq:VeffSMdS}) into
\begin{eqnarray}
V_{\rm H}(h,\mu, R)=\f{\xi(\mu)}{2}Rh^2+\f{\lambda(\mu)}{4}h^4+\frac{\alpha(\mu)}{144} R^2 + \Delta V_{\rm loops} (h, \mu, R)  \, .
\label{eq:Higgs-Potential}
\end{eqnarray}

In this result $\mu$ is an arbitrary dimensionful constant. The convergence of the perturbative expansion depends on the chosen value, and in general there is no single choice that gives good convergence for all values of $h$.
However, one can obtain an expression that is a good approximation of all $h$ using Renormalization Group Improvement (RGI), which sets the renormalization scale $\mu=\mu_*(h,R)$ as such that the loop correction to the potential vanishes~\cite{Ford:1992mv}, 
\begin{equation}
\Delta V_{\rm loops} (h, \mu_*, R) = 0.
\label{eq:dcond}
\end{equation}
Hence, we obtain the RG improved effective Higgs potential which reads
\begin{eqnarray}
    V_{\rm H}^{\rm RGI}(h, R) = \frac{\xi(\mu_*(h, R))}{2}  R h ^2 +\frac{\lambda(\mu_*(h, R))}{4}  h^4 + \frac{\alpha(\mu_*(h,R))}{144} R^2 \, ,
    \label{eq:RGIPot}
\end{eqnarray}
and which has no explicit loop correction term because of the  particular choice of the RG scale. Similarly to Ref.~\cite{Mantziris:2020rzh}, in the expressions above $h$ is referring to the renormalized Higgs field at the scale $\mu_*$, which is related to the renormalized field at some fixed physical scale $\mu_0$ through the anomalous dimension $\gamma$.

\begin{figure}[ht!]
    \centering
    \includegraphics[scale=0.8]{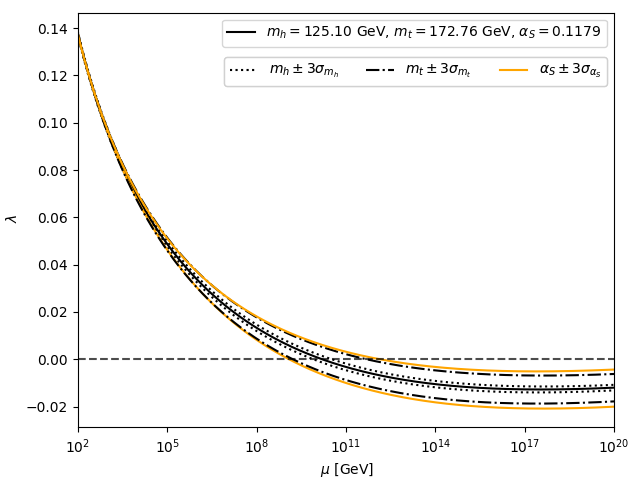}
    \caption{Running of the Higgs self-interaction coupling with the renormalization scale for the central values of the key parameters and $3 \sigma$ deviation around them.}
     \label{fig:lambda_running}
\end{figure}

The running couplings in Eq.~(\ref{eq:RGIPot}) are obtained by solving the renormalization group equations. For this we use the three-loop Minkowski space beta functions calculated in Ref.~\cite{Chetyrkin:2012rz}. Vacuum metastability arises because the Higgs self-coupling $\lambda(\mu)$ becomes negative at $\mu\gtrsim 10^{10}~{\rm GeV}$, as shown in Figure \ref{fig:lambda_running}. Because the renormalization scale $\mu_*$ is approximately equal to the largest of $h$ and $R$, it follows that when $R\lesssim 10^{10}~{\rm GeV}$, the RGI effective potential (\ref{eq:RGIPot}) has a barrier at $h\sim 10^{10}~{\rm GeV}$ making the vacuum metastable, and when $R\gtrsim 10^{10}~{\rm GeV}$ the barrier disappears altogether making the vacuum unstable unless it is stabilised by a sufficiently large and positive non-minimal coupling term $\xi$~\cite{Herranen:2015ima,Markkanen:2018bfx}.

\subsection{Effective masses in Einstein frame} \label{sec:effective_masses}

The previous discussion and results are applicable in an almost identical manner in the $R+R^2$ gravity scenario. Again, we assume a constant spacetime curvature $R$ and also a constant inflaton field $\phi$. This means that $\partial_\mu\phi=0$, and also that we can eliminate the inflaton field $\phi$ from the equations using the relation
\begin{equation}
    e^{-\sqrt{\frac{2}{3}}\frac{\phi}{M_P}} = 1 - \frac{2 H}{M M_P} = 1 - \frac{2 \sqrt{R/12}}{M M_P},
\end{equation}
which follows from the Friedmann equation.

Considering first the Higgs field, we can see from Eq.~(\ref{eq:Utilde}), that within our approximations, the relevant part of the action has the same form as in Einstein gravity, apart from a modified mass term. In terms of the rescaled mean field
\begin{align}
    \tilde{h} =  \Omega^{-1} h = e^{-\frac{1}{2} \sqrt{\frac{2}{3}}\frac{\phi}{M_P}} h \; , 
    \label{eq:hCT}
\end{align}
the effective masses of the Higgs and Goldstone modes are
\begin{align}
    \Tilde{m}^2_h &= m^2 e^{-\sqrt{\frac{2}{3}}\frac{\phi}{M_P}} +3\lambda \Tilde{h}^2 \, , 
    \label{eq:m_h} \\
    \Tilde{m}^2_\chi &= m^2 e^{-\sqrt{\frac{2}{3}}\frac{\phi}{M_P}}+\lambda\Tilde{h}^2 \, .
    \label{eq:m_chi}
\end{align}
In a similar way, all Standard Model fields can be rescaled in such as way that the form of the quadratic terms in the action is identical to Einstein gravity, but with potentially modified mass terms, denoted with a tilde ($\sim$).

For any fermion $\psi$ we therefore define a rescaled field $\tilde\psi$,  and a rescaled mass $\tilde{m}^2_\psi$ by demanding that
\begin{align}
    S_{\psi} = \int d^4x \sqrt{-g_J} \left ( i \Bar{\psi} \hat{\mathcal{D}} \psi - m_{\psi}\Bar{\psi} \psi \right) = \int d^4x \sqrt{-g} \left ( i \Bar{\Tilde{\psi}} \hat{\Tilde{\mathcal{D}}} \Tilde{\psi} - \Tilde{m}_{\psi}\Bar{\Tilde{\psi}} \Tilde{\psi} \right) \, ,
\end{align}
where the covariant derivative $\hat{\tilde{\cal D}}$ transforms under the conformal transformation as~\cite{Gorbunov:2010bn}
\begin{equation}
        \hat{\mathcal{D}} = \Omega \hat{\Tilde{\mathcal{D}}} = e^{\frac{1}{\sqrt{6}}\frac{\phi}{M_P}} \hat{\Tilde{\mathcal{D}}} \, . 
\end{equation}
This implies
\begin{align}
    \psi = \left( \Omega^2\right)^{3/4} \Tilde{\psi} = e^{\frac{3}{4} \sqrt{\frac{2}{3}}\frac{\phi}{M_P}} \Tilde{\psi} \, , \\ 
    \Tilde{m}^2_{\psi} = \f{y_{\psi}^2}{2}\Tilde{h}^2 \, .
    \label{eq:fermionsCT}
\end{align}

For $W$-bosons (and identically for $Z$-bosons), with gauge fixings $\zeta_i =1$, there is no field redefinition as $ \left( W^J \right) ^+_{\mu} = W^+_{\mu} $ since the exponential factors coming from the conformal transformation are cancelling one another in the kinetic term of the action
\begin{multline}
    S_{W} = \int d^4x \sqrt{-g_J} \left ( - g_J^{\mu \nu} g_J^{\sigma \rho} \partial_{\sigma} W_{\mu}^{+} \partial_{\rho} W_{\nu}^{-}  + m^2_{W} g_J^{\mu \nu} W^+_{\mu} W^-_{\nu} + W^+_{\mu} R_J^{\mu \nu} W^-_{\nu} \right) \\
    =  \int d^4x \sqrt{-g} \left ( - g^{\mu \nu} g^{\sigma \rho} \partial_{\sigma} W_{\mu}^{+} \partial_{\rho} W_{\nu}^{-}  + \Tilde{m}^2_{W} g^{\mu \nu} W^+_{\mu} W^-_{\nu} + W^+_{\mu} \left(R^{\mu \nu} + ... \right) W^-_{\nu} \right) \, ,
\end{multline}
where the dots ($\ldots$) indicate terms that vanish because we are assuming constant $\phi$. Therefore, this has the same form as in Einstein gravity, with the masses given in terms of the transformed Higgs field $\tilde{h}$ by 
 \begin{align}
\Tilde{m}^2_W &= \frac{g^2}{4}\Tilde{h}^2 \,,  
\label{eq:mtildeW} \\
\Tilde{m}^2_Z &= \frac{g^2 + (g')^2}{4}\Tilde{h}^2 \,.
\label{eq:mtildeZ}
\end{align}

From Eqs.~(\ref{eq:fermionsCT}), (\ref{eq:mtildeW}) and (\ref{eq:mtildeZ}) we can see that the particle masses have their standard expressions when written in terms of the transformed Higgs field, with the exception of the Higgs and the Goldstone bosons that receive an exponential suppression to their constant term as seen in Eqs.~(\ref{eq:m_h}) and (\ref{eq:m_chi}).

\begin{table}
	\caption{\label{tab:contributions}{Loop corrections to the effective potential with tree-level couplings to the Higgs, from the $\Tilde{W}^{\pm}$ and $\Tilde{Z^0}$ bosons, the quarks $\Tilde{\rm q}$, the leptons $\Tilde{l}$, the Higgs $\Tilde{h}$, the Goldstone bosons $\Tilde{\chi}_{W}$ and $\Tilde{\chi}_{Z}$, and the ghosts $\Tilde{c}_{W}$ and $\Tilde{c}_{Z}$, and corrections that do not to couple to the Higgs at tree-level, from the photon $\Tilde{\gamma}$, the gluons $\Tilde{g}$, the neutrinos $\Tilde{\nu}$, and the ghosts $\Tilde{c}_\gamma$ and $\Tilde{c}_g$.}}
	\vspace{2mm}
	\begin{center}
		{\tabulinesep=1.6mm
			\begin{tabu}{|c||ccccc|}
				\hline
				$\Tilde{\Psi}$ & $~~i$ & $~~n_i$   & $~d_i$    &$~n'_i$      &$\hspace{-2.6cm} \Tilde{\mathcal{M}}_i^2$    \\\hhline{|=#=====|}
				$~$ & $~~1$  & $~~2$       & $\quad{3}/{2}~~$        & $-34/15$        &  $\hspace{-2.2cm}\Tilde{m}^2_W+ R/12$      \\
				$~\Tilde{W}^\pm$ & $~~2$  & $~~6$       & $\quad{5}/{6}~~$       & $-34/5$        &  $\hspace{-2.2cm}\Tilde{m}^2_W+R/12$       \\
				$~$ & $~~3$  & $-2$      & $\quad{3}/{2}~~$         & $~~4/15$        & $\hspace{-2.4cm} \Tilde{m}^2_W-R/6$        \\\hline
				$~$ & $~~4$  & $~~1$        & $\quad{3}/{2}~~$ & $-17/15$        & $\hspace{-2.3cm} \Tilde{m}^2_Z+R/12$      \\
				$\Tilde{Z}^0$ & $~~5$  & $~~3$        & $\quad{5}/{6}~~$ & $-17/5$        & $\hspace{-2.3cm} \Tilde{m}^2_Z+R/12$     \\
				$~$ & $~~6$  & $-1$      & $\quad{3}/{2}~~$ & $~~2/15$        & $\hspace{-2.5cm} \Tilde{m}^2_Z-R/6$     \\\hline
				$\Tilde{\rm q}$ & $7-12$  & $-12$     & $\quad{3}/{2}~~$    & $~~38/5$        & $\hspace{-2.4cm} \Tilde{m}^2_{q}+R/12$      \\\hline
				$\Tilde{l}$ & $13-15$  & $-4$     & $\quad{3}/{2}~~$    & $~38/15$        & $\hspace{-2.4cm} \Tilde{m}^2_{l}+R/12$      \\\hline
				$\Tilde{h}$ & $~16$  & $~~1$       & $\quad{3}/{2}~~$          & $-2/15$      & $\hspace{-1.5cm} \Tilde{m}_h^2+(\xi-{1}/{6})R$  \\\hline
				${\Tilde{\chi}}_W$ & $~17$  & $~~2$      & $\quad{3}/{2}~~$           & $-4/15$      & $\hspace{0.2cm} \Tilde{m}_\chi^2\,+\zeta_W \Tilde{m}^2_W+(\xi-{1}/{6})R$   \\\hline
				${\Tilde{\chi}}_Z$ & $~18$  & $~~1$       & $\quad{3}/{2}~~$           & $-2/15$      & $\hspace{-0.1cm} \Tilde{m}_\chi^2+\zeta_Z \Tilde{m}^2_Z+(\xi-{1}/{6})R$   
				\\\hline ${\Tilde{c}}_W$ & $~19$  & $-2$       & $\quad{3}/{2}~~$           & $~~4/15$      & $\hspace{-2.0cm}\zeta_W \Tilde{m}^2_W-R/6$  
				\\\hline
				${\Tilde{c}}_Z$ & $~20$  & $-1$      & $\quad{3}/{2}~~$            & $~~2/15$      & $\hspace{-2.2cm}\zeta_Z \Tilde{m}^2_Z-R/6$   
				\\\hline
				$~$ & $~21$  & $~~1$       & $\quad{3}/{2}~~$        & $-17/15$        &  $ \hspace{-2.2cm} R/12$      \\
				$~\Tilde{\gamma}$ & $~22$  & $~~3$       & $\quad{5}/{6}~~$       & $-17/5$        &  $ \hspace{-2.2cm} R/12$       \\
				$~$ & $~23$  & $-1$      & $\quad{3}/{2}~~$         & $~~2/15$        & $ \hspace{-2.5cm} -R/6$        \\\hline
				$~$ & $~24$  & $~~8$        & $\quad{3}/{2}~~$ & $-136/15$        & $ \hspace{-2.2cm} R/12$      \\
				$\Tilde{g}$ & $~25$  & $~~24$        & $\quad{5}/{6}~~$ & $-136/5$        & $ \hspace{-2.2cm} R/12$     \\
				$~$ & $~26$  & $-8$      & $\quad{3}/{2}~~$ & $~~16/15$        & $\hspace{-2.5cm} -R/6$     \\\hline
				$\Tilde{\nu}$ & $27-29$  & $-2$      & $\quad{3}/{2}~~$           & $~~19/15$      & $\hspace{-2.2cm} R/12$   \\\hline
				${\Tilde{c}}_\gamma$ & $~30$  & $-1$       & $\quad{3}/{2}~~$           & $~~2/15$      & $\hspace{-2.5cm} -R/6$  
				\\\hline
				${\Tilde{c}}_g$ & $~31$  & $-8$      & $\quad{3}/{2}~~$            & $~~16/15$      & $\hspace{-2.5cm} -R/6$   
				\\\hline
			\end{tabu}
		}
	\end{center}
	
\end{table}

Because in these rescaled variables, the action has the same form as in Einstein gravity, the one-loop curvature correction to the effective potential is also identical to Eq.~(\ref{eq:DeltaVloops}) when expressed in terms of $\tilde{h}$, but with modified masses $\tilde{\cal M}_i$ given in Table \ref{tab:contributions},
\begin{align}
     \Delta V_{\rm loops} (\Tilde{h}, \mu_*, R) = \frac{1}{64\pi^2} \sum\limits_{i=1}^{31}\bigg\{ n_i \Tilde{\mathcal{M}}_i^4 \bigg[\log\left(\frac{|\Tilde{\mathcal{M}}_i^2 |}{\mu_*^2}\right) - d_i \bigg] +\frac{n'_i}{144} R^2 \log\left(\frac{|\Tilde{\mathcal{M}}_i^2 |}{\mu_*^2}\right)\bigg\} \,. \,\,\,\,\,\,
     \label{eq:DeltaVloops2}
\end{align}
In analogy with Eqs.~(\ref{eq:dcond}) and (\ref{eq:RGIPot}), the RGI improved effective potential is given by
\begin{eqnarray}
    V_{\rm H}^{\rm RGI}(\tilde{h}, R) = \frac{\xi(\mu_*(\tilde{h}, R))}{2}  R \tilde{h} ^2 +\frac{\lambda(\mu_*(\tilde{h}, R))}{4}  \tilde{h}^4 + \frac{\alpha(\mu_*(\tilde{h},R))}{144} R^2 \, ,
    \label{eq:RGIpot2}
\end{eqnarray}
where the renormalization scale $\mu_*(\tilde{h},R)$ is determined by demanding
\begin{equation}
\Delta V_{\rm loops} (\tilde{h}, \mu_*, R) = 0.
\label{eq:RGIcondition}
\end{equation}

Since we are assuming classical gravity, the inflaton field $\phi$ is also treated classical background field, and therefore neither it nor the graviton loops contribute to the beta functions. As a consequence of this, the beta functions used to obtain the running couplings in Eq.~(\ref{eq:RGIpot2}) are the standard ones, and not the ones shown in Ref.~\cite{Ema:2020evi}. This is a good approximation because the relevant energy scales, the highest of which is the Hubble rate during inflation $H_{\rm inf}\approx 10^{13}~{\rm GeV}$, are well below the Planck scale. Because of this, and because the masses $\Tilde{\mathcal{M}}_i$ appearing in the loops are almost identical to ${\mathcal{M}}_i$, the quantum corrections to the Higgs effective potential are very similar to those in Einstein gravity, apart from the rescaling of the field $h$.

\subsection{The effective potential in a time-dependent background} \label{sec:timedep}

While, as we found in the previous section, the quantum corrections are very similar to Einstein gravity, the extra classical terms in Eq.~(\ref{eq:action3}) play a very important role because of the time-dependence of the classical background. In order to incorporate this, we rewrite the action (\ref{eq:action3}) in a canonical form without neglecting the terms containing the inflaton field {and its derivatives}. We start by rewriting the Lagrangian (\ref{eq:action3}) more compactly as
\begin{align}
    \Lagr =  \frac{M_P^2}{2}R + \frac{A(\Tilde{h}, \mu_* )}{2}\partial_{\mu} \phi \partial^{\mu} \phi  + B(\Tilde{h}, \mu_* ) \partial_{\mu} \Tilde{h} \partial^{\mu} \phi + \frac{1}{2} \partial_{\mu} \Tilde{h} \partial^{\mu}  \Tilde{h}  - \Tilde{U} (\phi, \Tilde{h}, \mu_*) \, ,
     \label{eq:Lagrangian1}
\end{align}
where we have included the chosen RG scale explicitly and made the following definitions for compactness
\begin{align}
    A(\Tilde{h}, \mu_* ) &= 1 - \Xi(\mu_*) \left( \frac{ \Tilde{h} }{M_P} \right)^2  \, ,  \label{eq:factorA} \\
    B(\Tilde{h}, \mu_* ) &= - \sqrt{6} \Xi(\mu_*) \frac{\Tilde{h}}{M_P} \, ,
     \label{eq:factorB} \\
    \Xi(\mu_*) &= \xi(\mu_*)-\frac{1}{6} \, .
\end{align}
In the remainder of this section, we will be suppressing the $\mu_*$-dependence for clarity. 

In order to eliminate the mixing term, we perform the following field redefinition of the inflaton,
\begin{align}
    \phi = \Tilde{\phi} - M_P \sqrt{\frac{3}{2}} \mathrm{ln} \left[ 1 - \Xi \left( \frac{ \Tilde{h} }{M_P} \right)^2 \right] \, ,
    \label{eq:field-redef2}
\end{align}
which simplifies (\ref{eq:Lagrangian1}) into
\begin{align}
    \Lagr =  \frac{M_P^2}{2}R + \frac{A(\Tilde{h})}{2}\partial_{\mu} \Tilde{\phi} \partial^{\mu} \Tilde{\phi} + \frac{C(\Tilde{h})}{2} \partial_{\mu} \Tilde{h} \partial^{\mu}  \Tilde{h}   - \Tilde{U} (\Tilde{\phi}, \Tilde{h}) \, ,  \label{eq:Lagrangian3}
\end{align}
where we have again defined the following function for brevity
\begin{align}
    C(\Tilde{h} ) = 1 - \frac{ 6 \Xi^2\left( \frac{ \Tilde{h} }{M_P} \right)^2 }{1 - \Xi \left( \frac{ \Tilde{h} }{M_P} \right)^2 } \, .
    \label{eq:factorC}
\end{align}

We have to perform one last field redefinition, this time for the Higgs field, to bring the Lagrangian (\ref{eq:Lagrangian3}) into a canonical form
\begin{align}
        \Tilde{h} \approx \rho \left[ 1 + \Xi^2 \left( \frac{ \rho }{M_P} \right)^2 + \mathcal{O}(\rho^4) \right] \, .
        \label{eq:tildeh-to-rho}
\end{align}
Thus, finally we have an approximately diagonalised theory
\begin{align}
    \Lagr \approx  \frac{M_P^2}{2}R + \frac{1}{2}\partial_{\mu} \Tilde{\phi} \partial^{\mu} \Tilde{\phi} + \frac{1}{2} \partial_{\mu} \rho \partial^{\mu}  \rho   - \Tilde{U} (\Tilde{\phi}, \rho) \, ,  
    \label{eq:Lagrangian5}
\end{align}
where we have grouped all the potential terms in $\Tilde{U} (\Tilde{\phi}, \rho) = V_{\rm I} (\Tilde{\phi}) +  V_{\rm H}^{\rm RGI} (\Tilde{\phi}, \rho)$, with the first term corresponding to the Starobinsky potential (\ref{eq:V_I}) for $\Tilde{\phi}$ and the second to the RG improved effective Higgs potential 
\begin{align}
   V_{\rm H}^{\rm RGI} (\Tilde{\phi}, \rho ) = \frac{\alpha}{144} R^2 + \left[ \xi R +  \Delta m^2 \right] \frac{\rho^2}{2} + \left[ \lambda + \Delta \lambda_1 + \Delta \lambda_2 \right] \frac{\rho^4}{4} +  \frac{\Xi^2}{M_P^2} \left[ \lambda + \Delta \lambda_1 + \frac{\Delta \lambda_2}{8} \right] \rho^6 \, , \label{eq:V_H-potential3}
 \end{align}
with additional terms that were generated by the field redefinitions 
\begin{align}    
    \Delta m^2 &= 3 M^2 M_P^2 \Xi \left(1-e^{-\sqrt{\frac{2}{3}}\frac{\Tilde{\phi}}{M_P}}\right) e^{-\sqrt{\frac{2}{3}}\frac{\Tilde{\phi}}{M_P}} + \frac{\Xi }{M_P^2} \partial_{\mu} \Tilde{\phi} \partial^{\mu} \Tilde{\phi} \, , \\
    \Delta \lambda_1 &= 3 M^2\Xi^2 e^{-2\sqrt{\frac{2}{3}}\frac{\Tilde{\phi}}{M_P}} \, ,
    \label{eq:lambda-eff} \\
    \Delta \lambda_2 &= \frac{4 \Xi^2}{M_P^2} \left[ \xi R + 3 M^2 M_P^2 \Xi \left(1-e^{-\sqrt{\frac{2}{3}}\frac{\Tilde{\phi}}{M_P}}\right) e^{-\sqrt{\frac{2}{3}}\frac{\Tilde{\phi}}{M_P}} \right] + \frac{4 \Xi^3 }{M_P^4} \partial_{\mu} \Tilde{\phi} \partial^{\mu} \Tilde{\phi} \, .
\end{align}
Because we are interested in field values well below the Planck scale, the $\rho^6$ term is Planck suppressed, and we do not include it in the numerical calculations.

\section{Vacuum survival during inflation}

\subsection{Expected number of bubble nucleation events} \label{sec:2}

The experimental measurements of SM parameters, such as the Higgs boson and top quark masses, dictate that the EW vacuum state that the Higgs field resides in is metastable. This metastability originates from the sign switching of the Higgs self-coupling $\lambda$ as it runs with the renormalization scale $\mu$. If $\lambda(\mu > 10^{10} \, \mathrm{GeV}) < 0$, then the quartic potential of the Higgs field develops a potential barrier of finite height between the false and the true vacuum (whether bounded from below or not) which implies that it is possible for the Higgs field to go over or {tunnel} through the potential barrier, resulting in the formation of true-vacuum bubbles. These spherically symmetric vacuum bubbles grow rapidly with velocity close to the speed of light and within them spacetime is approximately anti de Sitter (AdS) and collapses to singularity. Therefore our existence implies that there can have been no bubble nucleation events in our past lightcone.

Denoting the probability that there were ${\cal N}$ bubbles in our past lightcone by ${\cal P}({\cal N})$, we therefore require ${\cal P}(0)\approx 1$, because otherwise our existence would be highly unlikely. Because we are interested in cases where bubble nucleation is extremely unlikely, we can assume that this probability distribution follows Poisson statistics. Then ${\cal P}(0)=\exp(-\langle {\cal N}\rangle)$, where $\langle{\cal N}\rangle$ is the expectation value of the number of bubbles in our past lightcone. The condition for vacuum stability can therefore be expressed as $\langle{\cal N}\rangle\lesssim 1$. This is convenient because if we know the cosmological history and can compute the nucleation rate per spacetime volume  $\Gamma(x)$ as a function of spacetime position $x$, then
\begin{align}
\Nbub = \int_{\rm past} d^4x \sqrt{-g} \Gamma(x) \, ,
\label{eq:expN0}
\end{align}
where the subscript ``past'' dictates that we are integrating over our past lightcone.

In this paper, we focus on the contribution from the period of inflation, which we denote by $\langle{\cal N}\rangle_{\rm inf}$. Because the integrand in Eq.~(\ref{eq:expN0}) is positive, the contribution from the rest of the cosmological history is positive, and therefore $\langle{\cal N}\rangle\ge\langle{\cal N}\rangle_{\rm inf}$. This means that if any inflationary scenario gives $\langle{\cal N}\rangle_{\rm inf}>1$, it is ruled out.

As the time coordinate to parameterise the inflationary era, we define the number of $e$-foldings $N=\ln (a_{\rm inf}/a)$, where $a$ is the scale factor and $a_{\rm inf}$ is a fixed reference time. This means that $N$ is counted backwards from the of inflation, and larger values correspond to earlier times during inflation.
We choose the reference point $a_{\rm inf}$ to correspond to the time when the Ricci scalar vanishes, $R=0$, after inflaton\footnote{This definition differs slightly from the one used in Ref.~\cite{Mantziris:2020rzh}, corresponding to shift by $\Delta N=0.192212$ in the definition of $N$. This shift is well below the accuracy of Eq.~(\ref{eq:ainf}).}. Defining the scale factor today to be $a_0=1$, this gives~\cite{Markkanen:2018pdo,Liddle:2003as}
\begin{equation}
\label{eq:ainf}
    a_{\rm inf}= \left( \frac{ H_0 e^{60}}{10^{16} \, \, \mathrm{GeV} } \right) \frac{V_{\rm I}^{1/4}(\phi_{\mathrm{inf}})}{H_{\mathrm{inf}}} \, .
\end{equation}
For Starobinsky inflation, $H_{\rm inf} \approx 6.5 \times 10^{12}$ GeV and $\phi_{\rm inf} \approx  0.6 M_P$, and using $H_0 \approx  1.5 \times 10^{-42}$ GeV, we obtain $ a_{\rm inf}\approx 1.25 \times 10^{-29}$.

We do not assume that the slow-roll conditions are necessarily satisfied, but we assume that the energy density is dominated by a homogeneous inflaton field $\phi$. Then the Hubble rate is given by
\begin{align}
H^2 =  \frac{V_{\rm I}(\Tilde{\phi})}{3M_P^2} \left[ 1 -  \frac{1}{6 M_P^2} \left( \frac{d \Tilde{\phi}}{d N} \right)^2 \right]^{-1} \, ,
\label{eq:Hubble-N}
\end{align}
and the Ricci scalar as
\begin{align}
    R = 12H^2 \left[ 1 -  \frac{1}{4 M_P^2} \left( \frac{d \Tilde{\phi}}{d N} \right)^2 \right] \, .
    \label{eq:Ricci-N}
\end{align}

In terms of conformal time $\eta$, defined as $dt=ad\eta$ with $\eta=0$ at the end of inflation, the comoving  radius of the past lightcone at conformal time $\eta$
is $r(\eta)=\eta_0-\eta$, where $\eta_0\approx 3.21/H_0$ is the conformal time today~\cite{Markkanen:2018pdo}. Therefore we can write Eq.~(\ref{eq:expN0}) as~\cite{Markkanen:2018pdo}
\begin{align}
    { \Nbub_{\rm inf} =  \int_{N_{\mathrm end}}^{N_{\mathrm{start}}} dN 
    \frac{4\pi}{3 H(N)}\left( \frac{a_{\mathrm{inf}} \left[\eta_0-\eta\left(N\right)\right]}{e^{N}} \right)^3 \Gamma(N) } \, ,
    \label{eq:Nbub-N}
\end{align}
where the limits of the integration $N_{\rm start}$ and $N_{\rm end}$ correspond roughly to the start and end of inflation, respectively, but we will discuss them in detail shortly.

In de Sitter spacetime with constant Ricci scalar $R$\footnote{With a dynamical metric, as done in Ref.~\cite{Hawking:1981fz}, the result is slightly different due to gravitational backreaction, but because in our case, the relevant energy scales are well below the Planck scale, the difference is minimal.} and constant inflaton field $\tilde{\phi}$, when $R$ is sufficiently high, the decay rate is given by the Hawking-Moss rate~\cite{Hawking:1981fz,Rajantie:2017ajw}
\begin{align}
    \Gamma_{\rm HM}( \Tilde{\phi}, R) &\approx \left(\f{R}{12}\right)^2 e^{-B_{\rm HM}( \Tilde{\phi}, R)}  \, ,
    \label{eq:Gamma} \\
    B_{\rm HM}( \Tilde{\phi}, R) &= \frac{384 \pi^2 \Delta V^{\rm RGI}_{\rm H} ( \Tilde{\phi}, R) }{R^2}  \, ,
    \label{eq:BHM}
\end{align}
where $\Delta V^{\rm RGI}_{\rm H} ( \Tilde{\phi}, {R}) = V^{\rm RGI}_{\rm H}(\rho_{\rm bar}, \Tilde{\phi}, R) - V^{\rm RGI}_{\rm H}(\rho_{\rm fv}, \Tilde{\phi}, {R})$ is the potential barrier height from the top of the barrier $\rho_{\rm bar}$ to the false vacuum $\rho_{\rm fv}$ ~\cite{Markkanen:2018pdo, Hawking:1981fz, Linde:1998gs}. 
We use Eq.~(\ref{eq:Gamma}) to approximate the decay rate in the time-dependent inflationary spacetime by replacing $R$ and $\tilde{\phi}$ by their time-dependent values,
\begin{equation}
    \Gamma(N)\approx \Gamma_{\rm HM}\left(\tilde{\phi}(N),R(N)\right) \, .
    \label{eq:GammaN}
\end{equation}

The choice of the limits of integration in Eq.~(\ref{eq:Nbub-N}), $N_{\rm start}$ and $N_{\rm end}$, is a compromise between stronger and more reliable bounds.
Specifically, they have to be chosen in such a way that the approximation (\ref{eq:GammaN}) we use when calculating the decay rate $\Gamma(N)$ is valid throughout. This is only the case when the spacetime can be well approximated by de Sitter spacetime. Deviation from de Sitter can be characterised by the adiabaticity parameter $\dot{H}/H^2$, which would be equal to zero in de Sitter. One can therefore expect Eq.~(\ref{eq:GammaN}) to be valid when
\begin{equation}
    \left|\frac{\dot{H}}{H^2}\right|\ll 1 \, .
    \label{eq:validity}
\end{equation}

As $N\rightarrow \infty$, $\dot{H}/H^2\rightarrow 0$ monotonically from below. Therefore the further back in time we go, the better the de Sitter approximation (\ref{eq:GammaN}) becomes. This means that Eq.~(\ref{eq:validity}) does not constrain the upper limit $N_{\rm start}$, and we could even consistently choose $N_{\rm start}=\infty$. On the other hand, empirically we have only evidence for roughly 60 $e$-foldings of inflation, somewhat dependent on the post-inflationary evolution.

Conversely, the further forward we go in time, the more it deviates from zero. At $N=0$, where $R=0$, $\dot{H}/H^2=-2$, and when $\ddot{a}/a=0$, which corresponds to $N\approx 0.19$ and is often defined to be the end of inflation, $\dot{H}/H^2=-1$. Therefore, in order to ensure that the condition (\ref{eq:validity}) is satisfied, $N_{\rm end}$ needs to be sufficiently large, $N_{\rm end}\gtrsim O(1)$. We will parameterise the choice of $N_{\rm end}$ by the corresponding value of the adiabaticity parameter $\dot{H}/H^2$, which lies in the range
\begin{align}
    -2 \leq \frac{\dot{H}}{H^2} < 0 \, .
\end{align}

For the numerical evaluation of the integral~(\ref{eq:Nbub-N}), it is convenient to express it as a system of coupled differential equations, which we solved with Mathematica in the same manner as in~\cite{Mantziris:2020rzh}\footnote{Unfortunately, there was a typo in the differential equation (3.12) of Ref.~\cite{Mantziris:2020rzh} for $\frac{d^2\phi}{dN^2}$, where the inflationary potential in the numerator was incorrectly squared.},
\begin{align}
  \frac{d\Nbub}{dN} &= \gamma(N)= \frac{4\pi}{3} \left[a_{\mathrm{inf}} \left(\frac{3.21e^{-N}}{H_0} -\Tilde{\eta}(N)\right)\right]^3 \frac{\Gamma(N)}{H(N)}  \, ,
  \label{eq:dNbubs-dN}\\
  \frac{d \Tilde{\eta}}{dN} &= - \Tilde{\eta}(N) - \frac{1}{a_{\mathrm{inf}} H(N)} \,,
    \label{eq:detatilda-dN}\\
\frac{d^2\Tilde{\phi}}{dN^2} &=\frac{V_{\rm I}(\Tilde{\phi})}{M_P^2H^2}
\left(\frac{d\Tilde{\phi}}{dN}-M_P^2\frac{V_{\rm I}'(\Tilde{\phi})}{V_{\rm I}(\Tilde{\phi})}\right) \, ,
\label{eq:dphi-dN}
\end{align}
where $\tilde{\eta}=e^{-N}\eta$. The boundary conditions for the field $\tilde{\phi}$ and its derivative were set at $\tilde{\phi}= 20 M_P$ by demanding that the right-hand-side of Eq.~(\ref{eq:dphi-dN}) vanishes, which corresponds to the slow-roll approximation. Eq.~(\ref{eq:dphi-dN}) was then evolved forward in time to find the point at which the Ricci scalar (\ref{eq:Ricci-N}) vanishes, $R=0$, which defines the origin, $N=0$. The full set of Eqs.~(\ref{eq:dNbubs-dN})--(\ref{eq:dphi-dN}) was then evolved towards larger $N$, with the additional boundary conditions
\begin{eqnarray}
\Nbub(0) &=& 0 \, , \nonumber\\
\tilde\eta(0)&=& 0 \, .
\end{eqnarray}

\subsection{Bubble nucleation in the Standard Model}

Let us now apply the general framework introduced in Section~\ref{sec:2} to the case of the Standard Model, with the particular aim of constraining the value of the Higgs non-minimal coupling $\xi$. Whenever we refer to numerical values of $\xi$, we mean the $\overline{\rm MS}$ renormalized parameter at scale $\mu=m_t$, which we denote by $\xi_{\rm EW}$ for clarity. In Ref.~\cite{Mantziris:2020rzh}, we carried out the same analysis for the field theory inflation model with the same inflaton potential as in the current case, and obtained the bound
\begin{equation}
    \xi_{\rm EW}\gtrsim 0.059^{+0.007}_{-0.009} \, ,
\end{equation}
to which we will compare our findings.

For the calculation of the decay rate (\ref{eq:Gamma}), we use the RGI effective potential (\ref{eq:V_H-potential3}) without the Planck-suppressed sixth-order term,
\begin{align}
    V_{\rm H}^{\rm RGI} (\rho, N) = \alpha (\mu_*) {\frac{R^2(N)}{144}} + m^2_{\rm eff} (\mu_*, N) \frac{\rho^2}{2} + \lambda_{\rm eff} (\mu_*, N) \frac{\rho^4}{4} \, ,  \label{eq:V_H-potential4}
\end{align}
where
\begin{align}
     m^2_{\rm eff} = \xi (\mu_*) R(N) + 3 M^2 M_P^2 \Xi (\mu_*) \left(1-e^{-\sqrt{\frac{2}{3}}\frac{\Tilde{\phi}(N)}{M_P}}\right) e^{-\sqrt{\frac{2}{3}}\frac{\Tilde{\phi}(N)}{M_P}} + \frac{\Xi (\mu_*) H^2(N)}{M_P^2} \left( \frac{d \Tilde{\phi}}{d N} \right)^2 \, ,
     \label{eq:meff}
\end{align}
\begin{multline}
        \lambda_{\rm eff} = \lambda (\mu_*) + 3 M^2\Xi^2 (\mu_*) e^{-2\sqrt{\frac{2}{3}}\frac{\Tilde{\phi}(N)}{M_P}} +
    \frac{4 \Xi^3 (\mu_*) H^2(N)}{M_P^4} \left( \frac{d \Tilde{\phi}}{d N} \right)^2 \\ +  
        \frac{4 \Xi^2 (\mu_*) }{M_P^2} \left[\xi (\mu_*) R(N) + 3 M^2 M_P^2 \Xi (\mu_*) \left(1-e^{-\sqrt{\frac{2}{3}}\frac{\Tilde{\phi}(N)}{M_P}}\right) e^{-\sqrt{\frac{2}{3}}\frac{\Tilde{\phi}(N)}{M_P}} \right]   \, ,
\end{multline}
and where $\mu_*$ also contains implicit dependence on $\rho$ and $N$ through Eq.~(\ref{eq:RGIcondition}). Therefore the potential is not a polynomial. Nevertheless, to understand the shape of the potential, it is instructive to think of $m^2_{\rm eff}$ and $\lambda_{\rm eff}$ as the coefficients of the quadratic and quartic terms, respectively, and consider their dependence on $N$.

\begin{figure}[t]
    \centering
    \includegraphics[scale=0.6]{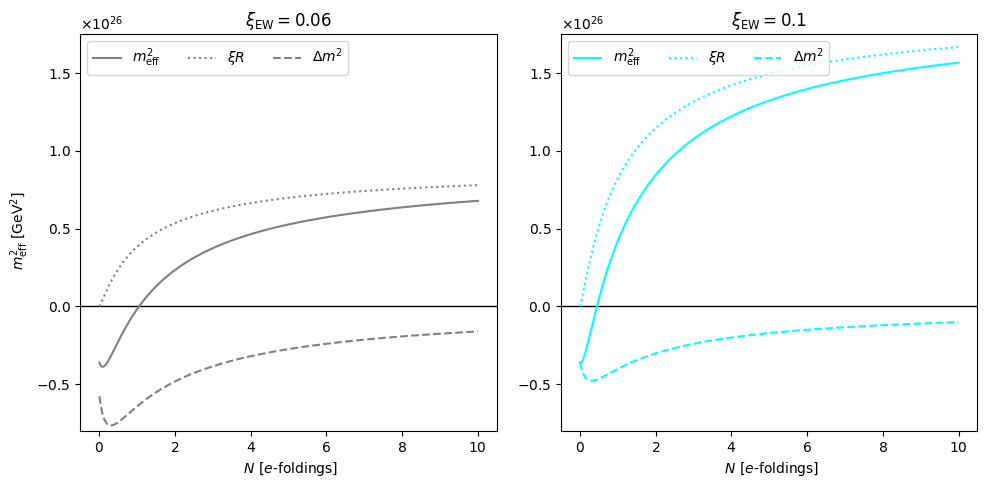}
    \caption{{Coefficient $m^2_{\rm eff}$} of the quadratic term in Eq.(\ref{eq:meff}) for $\xi_{\rm EW}=0.06$ (left) and $\xi_{\rm EW}=0.1$ (right), calculated at Higgs field value $\rho = 10^{12}$ GeV.}
    \label{fig:meff}
\end{figure}

\begin{figure}[h]
    \centering
    \includegraphics[scale=0.8]{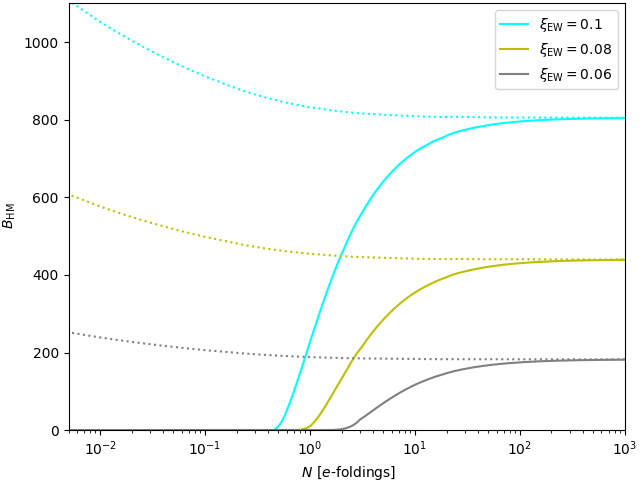}
    \caption{Bounce action {(\ref{eq:BHM})} for sample values of the non-minimal coupling $\xi_{\rm EW}$, during $R^2$ inflation (solid) and in comparison with the field theory case (dotted).}
    \label{fig:BHM}
\end{figure}

In Fig.~\ref{fig:meff}, we show the $N$-dependence of the coefficient $m^2_{\rm eff}$ of the quadratic term for two different values of $\xi_{\rm EW}$.
We can see that the extra contribution $\Delta m^2$, which is not present in the field theory inflation model considered in Ref.~\cite{Mantziris:2020rzh}, is negative and dominate over the non-minimal coupling term $\xi R$ at low $N$, very close to the end of inflation, destabilising the potential. This can also be seen more concretely in Fig.~\ref{fig:BHM}, which shows that value of the Hawking-Moss action (\ref{eq:BHM}) as a function of $N$ and compares it to the field theory inflation model, shown with dotted lines. Because the Hawking-Moss action is lower than in the field theory model, the decay rate (\ref{eq:Gamma}) is higher, and therefore vacuum stability will require a higher value of  $\xi_{\rm EW}$. The vanishing action at low $N$ indicates unsuppressed bubble nucleation, but only if the validity condition (\ref{eq:validity}) is satisfied, which means that the bound we obtain on $\xi_{\rm EW}$ are going to depend on the choice of $N_{\rm end}$.

Early on during inflation, when $N\gg 1$, the slow-roll approximations hold and spacetime looks approximately dS with the Hubble rate tending to a constant value $H \rightarrow H_{\rm dS} = \frac{M M_P}{2}$. The extra terms  in  the effective potential (\ref{eq:V_H-potential4}) are negligible, and therefore we find identical behaviour to Ref.~\cite{Mantziris:2020rzh}. This can also be seen by the behaviour of the bounce action in Fig. \ref{fig:BHM} for high values of $N$, where the two different curves overlap {for each $\xi_{\rm EW}$}.

\begin{figure}[h]
    \centering
    \includegraphics[scale=0.8]{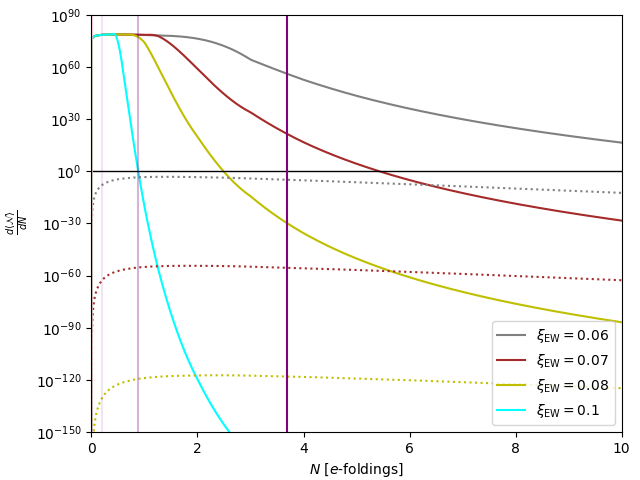}
    \caption{Integrands of $\Nbub$ with varying definition for the end of inflation. The vertical lines are at $\frac{\Dot{H}}{H^2} = -1, \, -\frac{1}{4} , \, -\frac{1}{32}$ respectively, and the dotted lines correspond to the field theory inflation model discussed in Ref.~\cite{Mantziris:2020rzh}. The plateau at $d\Nbub/dN\sim 10^{80}$, which the curves reach at small $N$, corresponds to vanishing Hawking-Moss action (\ref{eq:BHM}). In that case the expression (\ref{eq:Gamma}) for the bubble nucleation rate $\Gamma$  is not valid quantitatively, so the numerical value should be taken to be indicative of unsuppressed bubble nucleation, }
    \label{fig:integrands}
\end{figure}

This comparison between the field theory example of Ref.~\cite{Mantziris:2020rzh} and the proper implementation of $R^2$ inflation can be also made at the full integrand level, as shown in Fig. \ref{fig:integrands}. Once again, each pair of curves {approach each other} at earlier times, but they show very different evolution towards the final moments of inflation. The destabilising new terms in the potential lead to significantly higher integrands that result in an greater expectation number of true-vacuum bubbles, i.e. we have an enhancement of vacuum decay at late times. This effect is very sharply localised close to the very end of inflation meaning that bubble nucleation takes place predominantly moments before the inflationary finale.  The vertical lines, shown in varying shades of purple, denote different choices of $N_{\rm end}$, as indicated in the caption.

On the other hand, because the integrand decreases rapidly as a function of $N$, the number of bubbles $\Nbub_{\rm inf}$ and hence the bounds on $\xi_{\rm EW}$ are almost independent of the choice of $N_{\rm start}$, unless $N_{\rm start}\gtrsim 10^{60}$ as discussed in Ref.~\cite{Mantziris:2020rzh}. In practice, we therefore we only need to integrate up to $N=5$ to obtain precise bounds.

\section{Results} \label{sec:5}
Finding $\Nbub_{\rm inf}$ by solving the system of differential equation (\ref{eq:dNbubs-dN})-(\ref{eq:dphi-dN}), and demanding that $\Nbub_{\rm inf}<1$, results in a lower bound on the non-minimal Higgs curvature coupling $\xi_{\rm EW}$. This calculation is sensitive to the input SM parameters, {and because} the uncertainty in the estimation of the mass of the top quark is by far the greatest,  we explore the parameter space around its central value, $m_t = 172.76 \pm 0.30$ GeV. For a detailed account of the input parameters we use in this calculation see Table 1 in \cite{Mantziris:2020rzh}. This computation is also dependent on the choice of $N_{\rm end}$.

\begin{figure}[h]
    \centering
    \includegraphics[scale=0.86]{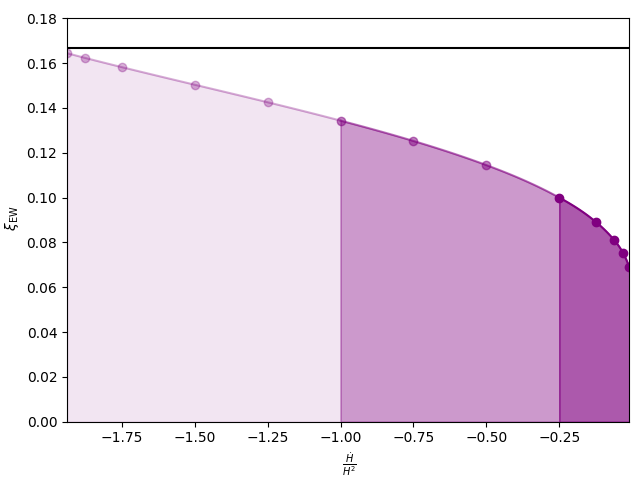}
    \caption{Dependence of the lower bound on the non-minimal Higgs curvature coupling $\xi_{\rm EW}$ on the choice of $N_{\rm end}$ in Eq.~(\ref{eq:Nbub-N}), parameterised by $\dot{H}/H^2$, for the top quark mass $m_t = 172.76$ GeV.The shaded regions below the curve denote the excluded values of the parameter space, the colour scheme ranges from the most conservative lower bounds in the darkest tone on the right to the less reliable in the lightest tone on the left and it matches with the corresponding bounds in figure \ref{fig:ksi-mtop}. The horizontal black line lies at the conformal value $\xi = 1/6$.}
     \label{fig:ksi-endinf}
\end{figure}

In Fig.~\ref{fig:ksi-endinf}, we present the effect of the choice of $N_{\rm end}$, which we parameterise by the adiabaticity parameter $\dot{H}/H^2$, on the lower $\xi$-bounds for the central value $m_t=172.76~{\rm GeV}$ of the top quark mass. A more negative $\dot{H}/H^2$ corresponds to lower $N_{\rm end}$ and therefore leads to a significantly stronger bound on $\xi_{\rm EW}$, approaching the conformal value at $\dot{H}/H^2=-2$. However, because they violate the validity condition (\ref{eq:validity}), these bounds probably cannot be trusted. However, the darkest purple area, corresponding to $\dot{H}/H^2>-1/4$, should be valid, and therefore we can conclude that vacuum stability requires $\xi_{\rm EW}\gtrsim 0.1$.

This result is in numerical agreement with Ref.~\cite{Li:2022wuf} which studied the same question recently, by considering only classical evolution rather than Hawking-Moss transitions. That calculation also does not include curvature corrections to the effective potential, apart from the tree-level non-minimal coupling term.

\begin{figure}[h]
    \centering
    \includegraphics[scale=0.86]{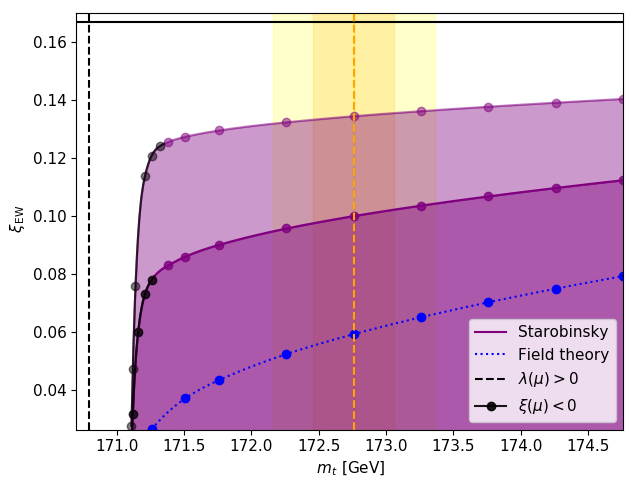}
    \caption{Lower bounds on the non-minimal Higgs curvature coupling $\xi_{\rm EW}$ as a function of the top quark mass $m_t$. The vertical dashed orange line with its accompanying shaded regions depict $m_t \pm \sigma \, , 2 \sigma$ \cite{Zyla:2020zbs}. The darker and lighter shades of purple show the excluded areas for two different choices of $N_{\rm end}$, corresponding to $\dot{H}/H^2=-1/4$ and $\dot{H}/H^2=-1$, respectively. The leftmost black parts of the curves show the lowest $\xi_{\rm EW}$ values below which $\xi(\mu)$ turns negative during its running, and thus ``pushes'' the EW vacuum to higher field values. Previous constraints with a Starobinsky-like power-law model are shown in the dotted blue curve \cite{Mantziris:2020rzh}. The vertical dashed black line stands at the threshold value of $m_t$, below which the EW vacuum is stable. Finally, the horizontal, black line illustrates the conformal point $\xi = 1/6$.}
    \label{fig:ksi-mtop}
\end{figure}

In figure \ref{fig:ksi-mtop}, we present the lower bounds with regard to the input value of the top quark mass. The shaded areas are excluded by vacuum instability. The darker and lighter shades of purple correspond to different choices of $N_{\rm end}$, $\dot{H}/H^2=-1/4$ and $\dot{H}/H^2=-1$, respectively.The blackened portions of the curves correspond to the area of parameter space where the non-minimal coupling turns negative as it runs and thus forces the metastable vacuum to higher field values close to the potential barrier. These bounds are obtained in a slightly different manner, since the height of the potential barrier is measured from the top of the barrier to the now dynamic local minimum of the potential. For comparison, we also show with the blue dotted curve the bounds for the field theory inflation model of Ref.~\cite{Mantziris:2020rzh}. They are evidently weaker, since now extra terms have been generated in the effective potential that have negative sign and therefore destabilise the vacuum increasingly towards the end of inflation.

\section{Conclusions}

We have studied electroweak vacuum stability in the minimal Starobinsky inflation model, in which inflation is driven by an $R^2$ term. In the Einstein frame, this term gives rise to a negative time-dependent contribution to the Higgs effective potential, whose effect is to destabilise the electroweak vacuum further.

We incorporated quantum effects in the Higgs effective potential by using the three-loop RGI effective potential together with one-loop curvature corrections computed in the de Sitter approximation with a constant inflaton field, and used this potential to compute the Hawking-Moss vacuum decay rate. By demanding that no bubble nucleation events took place in our past lightcone, we obtained a lower bound on the non-minimal Higgs curvature coupling $\xi_{\rm EW}$,
\begin{align*}
            \xi_{\rm EW} \gtrsim 0.1 \, .
        \end{align*}
This is significantly stronger than the corresponding bound $\xi_{\rm EW}\gtrsim 0.06$ obtained for the Starobinsky-like field theory inflation model~\cite{Mantziris:2020rzh}.

These constraints exhibited similar mild dependence on the top quark mass as in \cite{Mantziris:2020rzh}, but were significantly more sensitive to the last moments of inflation and therefore also to the precise choice of $N_{\rm end}$, the lower limit of the integral (\ref{eq:Nbub-N}). Because vacuum bubble production is pushed towards the end of inflation, where our de Sitter approximations start to break down, we adopted a conservative choice for $N_{\rm end}$, corresponding to the condition $\dot{H}/H^2=-1/4$. However, this suggests that the bounds may be improved significantly by fully accounting for the transition from inflation to radiation dominated hot Big Bang. Vacuum stability during reheating in this same theory was recently studied in Ref.~\cite{Li:2022wuf}, but further work is needed to bridge the gap between these two calculations.

On the other hand, when considering the shape of the effective potential at early times, we recover the results and the same cosmological implications from \cite{Mantziris:2020rzh}, where the bounds are not sensitive to the entire duration of inflation, unless it lasts for more than $10^{60}$ $e$-folds and thus eternal inflation appears to be inconsistent with vacuum metastability.

\section*{Acknowledgements}
AR was supported by STFC grants ST/P000762/1 and ST/T000791/1, and by an IPPP Associateship. AM was supported by an STFC PhD studentship.
This project has received funding from the European Union’s Horizon 2020 research and innovation programme under the Marie Sk\l odowska-Curie grant agreement No. 786564. TM also acknowledges support from Solita Oy's Connected Data and Research units.

\bibliography{references.bib}
\end{document}